\documentclass[largeformat]{interact}
\usepackage[colorlinks=true,allcolors=blue!80!black]{hyperref}

\usepackage[utf8]{inputenc}
\usepackage{amsmath}
\usepackage[caption = false]{subfig}
\usepackage{graphicx,epstopdf}
\usepackage{blindtext}
\usepackage{lipsum}
\usepackage{amsfonts}
\usepackage{bbm}
\usepackage{amssymb}
\usepackage{enumerate}
\usepackage{color}
\usepackage{latexsym}
\usepackage{physics}
\usepackage{times,txfonts}
\usepackage{xcolor}
\usepackage{epsfig}
\usepackage{bm}
\usepackage{tabularx}
\usepackage{multirow}
\usepackage{mathtools}
\usepackage{xfrac}
\usepackage{physics}
\usepackage{enumerate}
\usepackage{braket}
\usepackage{cite}
\usepackage{dsfont}
\usepackage{hyperref}

\begin{document}

\title{On the origin of force sensitivity in tests of quantum gravity \\ with delocalised mechanical systems}

\author{\name{Julen~S. Pedernales\thanks{Julen~S. Pedernales. Email: julen.pedernales@uni-ulm.de} and Martin~B. Plenio \thanks{Martin~B. Plenio. Email: martin.plenio@uni-ulm.de}\textsuperscript{a}} \affil{\textsuperscript{a} Institut f\"ur Theoretische Physik and IQST, Albert-Einstein-Allee 11, Universit\"at Ulm, D-89081 Ulm, Germany}}

\maketitle

\date{\today}
\begin{abstract}
The detection of the quantum nature of gravity in the low-energy limit hinges on achieving an unprecedented degree of force sensitivity with mechanical systems. Against this background we explore the relationship between the sensitivity of mechanical systems to external forces and properties of the quantum states they are prepared in. We establish that the main determinant of the force sensitivity in pure quantum states is their spatial delocalisation and we link the force sensitivity to the rate at which two mechanical systems become entangled under a quantum force.
We exemplify this at the hand of two commonly considered configurations. One that involves gravitationally interacting objects prepared in non-Gaussian states such as Schr\"odinger-cat states, where the generation of entanglement is typically ascribed to the accumulation of a dynamical phase between components in superposition experiencing varying gravitational potentials. The other prepares particles in Gaussian states that are strongly squeezed in momentum and delocalised in position where entanglement generation is attributed to accelerations. We offer a unified description of these two arrangements using the phase-space representation of the interacting particles, and link their entangling rate to their force sensitivity, showing that both configurations get entangled at the same rate provided that they are equally delocalised in space. Our description in phase space and the established relation between force sensitivity and entanglement sheds light on the intricacies of why the equivalence between these two configurations holds, something that is not always evident in the literature, due to the distinct physical and analytical methods employed to study each of them. Notably, our findings demonstrate that while the conventional computation of entanglement via the dynamical phase remains accurate for systems in Schr\"odinger-cat states, it may yield erroneous estimations for systems prepared in squeezed cat states. 
\end{abstract}

\begin{keywords}
{\it force sensing; weak forces; optomechanics; quantumness of gravity; gravitationally induced entanglement; matter-wave interferometry; macroscopic quantum superpositions.}
\end{keywords}

%
%
\section{Introduction}%

In science, precision matters. Improvements in the measurement precision of physical quantities often unveil new layers of reality and make them accessible for scientific scrutiny. For example, bacteria were discovered by the new tool of microscopy \cite{vanLeewenhoeck1677,WeisenburgerS2015}, X-ray diffraction imaged the structure of DNA \cite{FranklinG1953,Squires2003}, and gravitational waves were detected by optical interference experiments of unprecedented precision \cite{Abbott+2016,McIverS2021}. One of the most precise measurement techniques in the toolbox of modern science is, precisely, that of interferometry. In the broadest sense, interferometry refers to the use of interference to extract information from a system, usually, in the form of a relative phase between two entities that are made to interfere. For example, optical interferometers measure the phase difference between light beams travelling along different paths.  Ramsey interferometry measures the phase difference between the states of a two-level system in superposition, such as a spin that is in a superposition of being aligned and counter-aligned with respect to a static magnetic field. The phase then provides an accurate measurement of the intensity of the magnetic field. Similarly, the relative phase between two electronic states of an atom can be used to set ultra-precise frequency (and in turn time) standards \cite{EssenP1955,Margolis2010}. In fact, many physical magnitudes can be measured via interference by first mapping them onto the relative phase of a system in superposition. 

Of special interest to this article is the measurement of weak conservative forces. Consider a massive object, such as an atom, delocalised along two distinct trajectories. A force originating from a spatially inhomogeneous potential field imparts a different phase to each component of the superposition. The detection of such a relative phase is a measure of the gradient of the potential field and therefore of the force. On the other hand, such a force will also impart momentum to the massive object, something that could, in principle, be measured, and would also inform about the magnitude of the force. But, which of these two approaches shall provide a greater precision? It is, perhaps, tempting to state that interferometry is likely to provide better measurement precision, after all, interferometric approaches have proven to provide unprecedented precision across multiple systems and physical quantities. In this article, however, we will argue that both approaches offer the same fundamental sensitivity limit, provided that the massive object used for detection is delocalised in space to the same extent in both cases. But, why is the measurement precision of weak forces relevant? It has been argued that the detection of weak gravitational forces, not the gravitational forces generated by astronomical objects, but rather the forces exerted on each other by tiny nanometer-scale solids could shed light on the nature of the gravitational force itself \cite{DeWitt2011,Kibble1981,Peres2005,Taylor2013}. Let us see how that would be so.

There exists no accepted experiment or observation that demands the unification of general relativity and quantum mechanics. Yet, it is widely acknowledged that the current, separate treatment that these two theories receive constitutes an unsatisfactory 
framework for physics. The reason is simple: quantum mechanics claims that all matter is quantum, and general relativity that all matter distorts space-time geometry, albeit it only describes how classical matter does so. Since, in their current form, both theories are to govern all systems, in all regimes, a clash between the theories emerges in those regimes where both phenomenologies are relevant. 
For example, consider the gravitational interaction between two massive objects whose centres of mass are in a spatial quantum superposition. Which theory governs the evolution of the system? On the one hand, general relativity cannot describe such a configuration for its formalism does not make sense of sources of the gravitational field in superposition. On the other hand, within the framework of quantum mechanics, there is no known way of consistently describing the gravitational interaction across all energy regimes. The considered situation is, therefore, outside the realm of existing physical theories. But, at the same time, neither quantum mechanics nor general relativity prevent the imagined situation from occurring. It is a rather uncontroversial conclusion that either one of the two theories or both needs to be modified. 

The divergence of perspectives arises in the subsequent stage, that is when considering how to pursue a new theory that can handle quantum fields and gravity consistently. On the one hand, we find the program of {\it quantum gravity}, which challenges the fundamental character of general relativity and suggests that this should be modified in order to make it compatible with the principles of quantum mechanics. This is not a radical idea. In fact, it is a concept that surfaced quite early in the history of general relativity. In 1916, one year after publishing the general theory of relativity, Einstein famously wrote that ``{\it [i]t appears that quantum theory would have to modify not only Maxwellian electrodynamics but also the new theory of gravitation}''~\cite{Einstein2016}. At the time, however, quantum mechanics was not fully formalised and thus it was not until 1930 that the first technical efforts to quantise gravity took place by the hand of Rosenfeld~\cite{Rosenfeld1930, Rosenfeld1930b}. Interestingly, one year earlier, in their seminal paper on the quantisation of fields~\cite{Pauli1929}, Heisenberg and Pauli, had predicted that ``{\it [...] a quantisation of the gravitational field, which seems to be necessary on physical grounds, is also practicable by means of a formalism that is completely analogous to the one employed here with no new difficulties}''. However, this initial optimism was soon proven to be unfounded. Bronstein \cite{Bronstein1936} demonstrated that the quantisation of the electromagnetic field is fundamentally different from that of the gravitational field.
Over the ensuing decades, efforts to quantise the gravitational field have branched out in a myriad of distinct approaches, including prominent candidates such as string theory or loop quantum gravity~\cite{Rovelli2001,Palti2021}. While these efforts have yielded intriguing mathematical structures and promising results, the stark reality remains that, over a century after Einstein's initial suggestion, a comprehensive theory of quantum gravity that universally holds across all energy scales remains elusive.

Given the difficulty of quantising gravity, it has been suggested that, perhaps, the road should be travelled in the opposite direction. That is, perhaps, one should try to modify quantum mechanics to align it with the principles of general relativity, or as more provocatively put by Penrose, attempt the  ``gravitization"  of quantum mechanics~\cite{Penrose2014,Penrose2014b}.  Efforts following this route aim at finding a mechanism that can couple a classical gravitational field to quantum matter. By classical, here, we mean a field that has a definite value at each point in space. That is, a gravitational field that cannot enter a superposition. This can be the gravitational field described in standard general relativity or a modified version, where the field remains classical. Starting from the 1960s, several efforts have attempted to reconcile quantum mechanics and gravity without quantising the gravitational field~\cite{Moller1962, Rosenfeld1963, Karolyhazy1966, Diosi1987, Diosi1989, Penrose1996, Milburn2014, Diosi2016, Oppenheim2021, Oppenheim2022, Bassi2022}. Some of these have the added appeal that, through the same coupling to the gravitational field, they offer a mechanism for the collapse of the wave function, which could eventually describe the quantum-to-classical transition of the macroscopic world. Two fixes in one package. However, a fully consistent semiclassical gravity model has not yet been accomplished. In fact, it has been argued that the coupling between classical and quantum fields leads to a number of inconsistencies such as the violation of momentum conservation, the violation of the uncertainty principle, or faster than light signalling~\cite{Hannah1977, Kibble1981, Geilker1981, Unruh1984, Salcedo2012, Vedral2017}. Under these arguments, semiclassical models of gravity would only serve as approximations of an underlying quantum theory of gravity valid in some restricted regimes, but never as a fundamental theory. However, the validity of these claims has been repeatedly called into question~\cite{Oppenheim2021, Callender2001, Mattingly2006, Kent2018, Pikovski2021, Grossardt2022}, and the view that semiclassical theories of gravity are to be disregarded on theoretical grounds alone is by no means unanimous. The debate is open. 

It is noteworthy to mention that hybrid quantum-classical theories had also been considered for the case of the electromagnetic field~\cite{Slater1924, Jaynes1969}. However, a crucial difference sets them apart from their gravitational counterparts: these theories were abandoned due to compelling experimental proof of the quantisation of the electromagnetic field, including the detection of single photons~\cite{Geiger1924, Simon1925, Clauser1974, KimbleDM1977}. This is missing for the gravitational case. While it is likely safe to assert that a substantial consensus within the scientific community posits that gravity must ultimately be governed by quantum principles~\footnote{A first-order approximation of the sentiment within the physics community could, perhaps, be gleaned from a recent bet involving Jonathan Oppenheim, Carlo Rovelli and Geoff Penington, who set the odds $5000$ to $1$ in favor of gravity being quantum~\cite{Rovelli2022}}, the absence of concrete experimental corroboration justifies a stance of skeptical inquiry. In the words of Rosenfeld, ``{\it [t]here is no denying that, considering the universality of the quantum of action, it is very tempting to regard any classical theory as a limiting case to some quantal theory. In the absence of empirical evidence, however, this temptation should be resisted}"~\cite{Rosenfeld1963}.

This brings us to a pivotal query: Can we subject semiclassical theories of gravity to experimental testing, thus either confirming or refuting their validity? It is sometimes believed that experiments at the juncture of quantum mechanics and gravity are unfeasible, stemming either from the absence of testable predictions or the requirement of impractical levels of energy, mass, or scale~\cite{Dyson2013}. However, with the advent of controllable quantum technologies, and in particular optomechanical devices~\cite{Stickler2020,Romero-Isart2021a}, experimental access to the low-energy limit of the intersection between gravity and quantum mechanics might be closer than ever before. But, can experiments conducted within this low-energy regime indeed yield informative outcomes? The answer is yes. In particular, a Gedankenexperiment introduced by Feynman in the 1950s~\cite{DeWitt2011} is currently experiencing a resurgence of interest due to its potential realisation using contemporary quantum technologies. If attainable, this experiment holds the promise of serving as a test for semiclassical theories of gravity, akin to a Bell test tailored to gravitational theories. In such a scenario, a specific experimental outcome would be incompatible with semiclassical theories of gravity, thereby offering a means for their falsification. However, the implementation of such an experiment will require unprecedented precision in the measurement of the gravitational force between nano- or micro-scale objects, which brings us back to the question asked at the beginning of this article: shall this precision be pursued by leveraging the build-up of relative phases or by looking at the change in momentum of the interacting particles?

In the remainder of this paper, we will describe this Gedankenexperiment and its reformulation in the language of modern quantum technologies and quantum information theory. We will then present current experimental proposals for its implementation with optomechanical devices, dividing them into two categories: those that leverage relative phases and those that work under the change in momentum. Then we will offer a quantum optician's point of view of these experiments and argue that the two are fundamentally equivalent. This equivalence, though not necessarily at odds with the prevailing consensus, has sometimes sparked confusion due to the manner in which the two approaches are typically presented. In fact, the writing of the present article found its original motivation in the experience of one of the authors. When that author confidently stated the aforementioned equivalence without formal proof in a seminar, he found himself confronted with significant opposition expressing with equal confidence the opposing view. With this paper, we aim to dispel any confusion of the issues involved by offering a clear elucidation of how this equivalence is established---an endeavour that, to our knowledge, has not been explicitly undertaken thus far.

\section{Feynman's Gedankenexperiment and gravitationally induced entanglement}
In the winter of 1957, at Chapel Hill, North Carolina, the {\it Conference on the Role of Gravitation in Physics} took place, bringing together some of the most distinguished physicists of the time~\cite{Zeh2008,DeWitt2011}. The meeting often considered a pivotal event in the history of gravitation, is regarded as having ignited the search for gravitational waves and established quantum gravity as a field of its own~\cite{Rickles2021}. In a session discussing the necessity of gravitational quantisation, Feynmann presented a Gedankenexperiment aimed to illustrate the divergent experimental results that can be expected from treating gravity through either quantum or classical approaches. 

\begin{figure}[htbp]
\begin{center}
\includegraphics[width=0.81\textwidth]{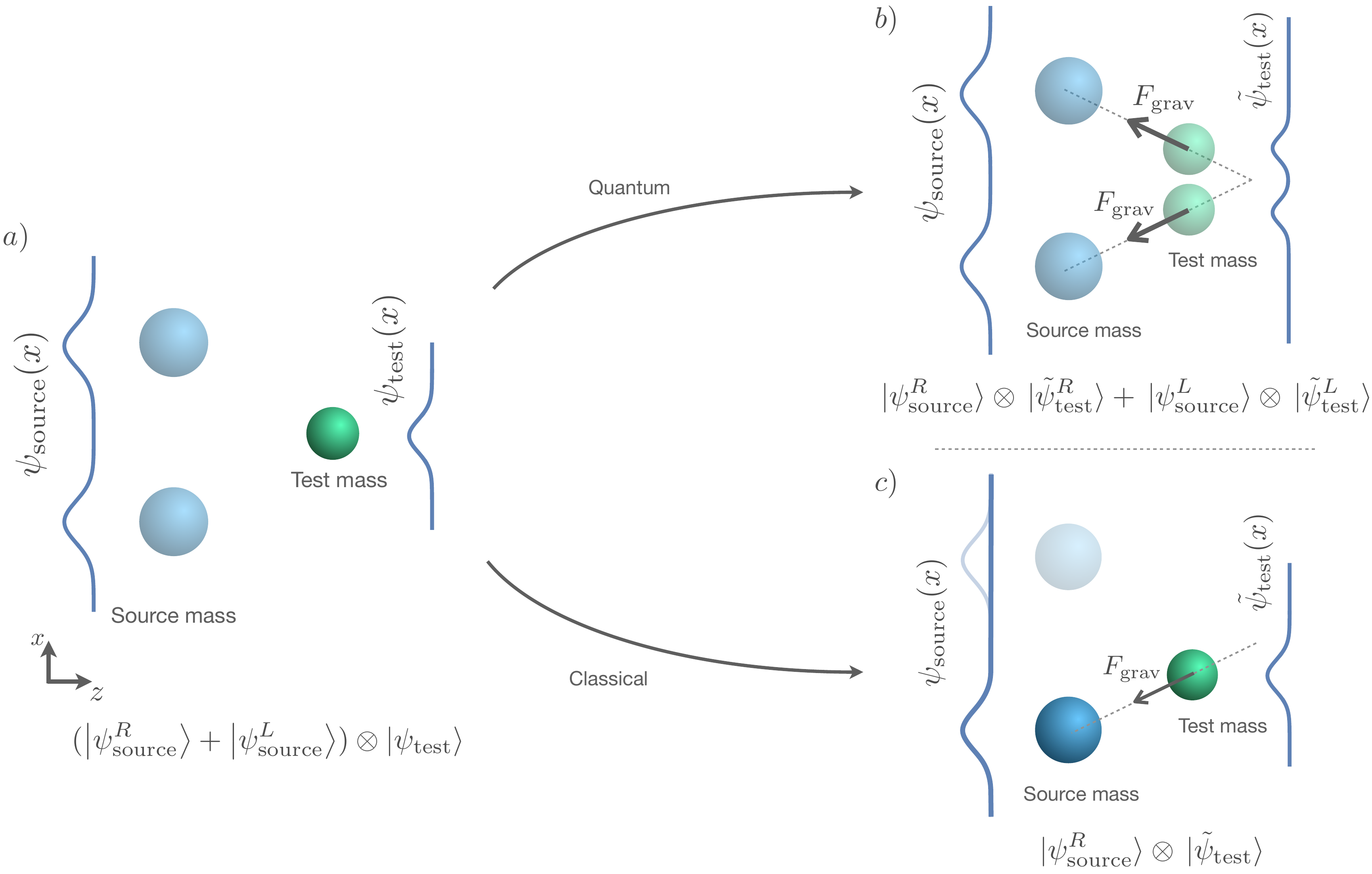}
\caption{{\bf Feynman's Gedankenexperiment.} a) A source of the gravitational field is placed in a quantum superposition (blue ball) and let to interact with a test mass (green ball). The system can evolve into two possible configurations: b) If the field is in a quantum superposition, the test mass experiences a superposition of two gravitational forces and consequently evolves into a superposition. c) If the field remains classical the test mass experiences a single force and moves consequently. For example, according to some semiclassical models, the source will collapse to one of its states in superposition, and the gravitational field will adopt a classical form corresponding to the source being located in that position. The test mass would then move accordingly, as shown pictorially in the drawing. For all three plots, the drawings of the wave functions indicate whether the objects are in a spatial superposition or not. These are not to be taken literally, notice, for example, that in b) the particles are entangled and their wave functions would not be expressable independently from each other.}
\label{Feynman}
\end{center}
\end{figure}

The logic of Feynman's idea is as follows: Take a sizeable massive object capable of generating a measurable gravitational field---the exact size, illustrated by Feynman with a 1-cm ball, is immaterial to the underlying argument. Next,  position the centre of mass of this object into a coherent superposition across two distinct locations in space, as depicted in Fig.~(\ref{Feynman}a). The practical realisation of this scenario is undeniably complex, but in the realm of a Gedankenexperiment, as long as it avoids fundamental impossibility, it remains valid~\footnote{To prepare the superposition of the ball, originally, Feynman describes a Stern-Gerlach-like apparatus where spin-1/2 particles are deflected to one of two counters. When a particle hits one of the counters it activates a mechanism that displaces the ball in a direction that is different for each counter. Provided that the mechanism that displaces the ball constitutes a quantum coherent interaction, this would prepare the ball in a superposition of two locations.}. The central query emerges: What gravitational field does the ball generate under these conditions?
 
To tackle this question, Feynman proposes introducing a second ball---which we will term the ``test mass"---and observing its reaction to the field. Assuming the interaction between the source and test masses is exclusively gravitational, two potential scenarios arise. If the gravitational field genuinely exists in a superposition state, the test mass would perceive two attractive forces, each directed toward the respective locations of the components of the source in superposition, as seen in Figure~\ref{Feynman}b. Consequently, the test mass would evolve into a superposition of its centre of mass, mirroring the behaviour of the gravitational field. However, if the gravitational field remains classical, the test mass would experience a single gravitational force, propelling its movement accordingly, without necessitating a division of its wavepacket.
Feynman points out that these two scenarios are fundamentally different, and can, in principle, be distinguished in an experiment. But, what exactly can we measure to distinguish them?

From the point of view of modern quantum information theory, it has been recognised that the fundamental, measurable difference between the two plausible experimental outcomes is the presence or absence of entanglement~\cite{Peres2005, Taylor2013, Aspelmeyer2016, Paterek(2017)}. It is not hard to convince ourselves that in scenarios where the gravitational field exhibits quantum behaviour, the positions of the source and test masses become entangled during their evolution under the influence of gravitational interaction, as illustrated in Figure~(\ref{Feynman}b). Conversely, when the gravitational field adheres to classical behaviour, any correlations between the positions of the two masses assume a classical character---manifesting as a statistical mixture. 

This notion, expressed here somewhat qualitatively, can be more rigorously formalised through a concept sometimes referred to as the LOCC theorem. Within the realm of quantum information theory, transformations of the state of a bipartite system achieved through either local operations (LO) on each partition separately, classical communication (CC) between the partitions, or a combination thereof are referred to as LOCC transformations. The LOCC theorem stipulates that this class of transformations cannot increase the degree of entanglement within the bipartite system it acts upon~\cite{VirmaniP07}.

In the context defined here, a semiclassical theory of gravity---where gravitational interactions are propagated by a classical field engaging in local interaction with quantum matter---produces, by definition, a state transformation within the LOCC framework. Consequently, the identification of entanglement between objects that initially exist in a separable state, and solely interact gravitationally, is fundamentally at odds with the premise of a semiclassical theory of gravity and would constitute a falsification of such theories. Following this line of reasoning, the notion of {\it gravitationally induced entanglement} has emerged as a test for the classical/quantum nature of gravity~\cite{Peres2005, Taylor2013, Aspelmeyer2016, Milburn2017, Vedral2017a, Paterek2020, Plenio2020, Plenio2021, Romero-Isart2021, Plenio2022} accompanied by a rich discussion on the precise extent of its power and limits as a test of quantum gravity~\cite{Rovelli2019, Aspelmeyer2018, Carney2022, DonerG2022, Plenio2022b, Wald2022, Howl2023, Rovelli2023, Plenio2023}.

It is worth noting that the possibility of non-local theories of gravity, where the gravitational interaction occurs not through a classical or quantum field but through direct action at a distance, has also been anticipated~\cite{Pikovski2022, Perche2023}. Such theories could, in principle, result in the generation of entanglement. Consequently, the identification of gravitationally induced entanglement would only serve to refute the proposition of a local, classical field theory of gravity, without necessarily serving as evidence for the quantisation of the gravitational field.

However, from our perspective, non-local theories of gravity would inherently constitute a quantum theory of gravity. Therefore, we hold the belief that the observation of gravitationally induced entanglement would indeed substantiate the quantum nature of gravity. Nonetheless, we acknowledge that this stance hinges on the underlying definitions and interpretations of the terms {\it quantum} and {\it classical}. 

Before closing this section, we would like to point out that other methods to find evidence against or in favour of semiclassical theories of gravity exists. These include, but are not restricted to, methods to reveal the coherence of gravitational fields in superposition~\cite{Peres2005, Bassi2017, Haine2019, Ulbricht2019}, the detection of the decoherence effects induced by semiclassical models~\cite{ Wehner2016, Weller-Davies2022}, the observation of non-Gaussian signatures of gravity~\cite{Iyer2021} or the violation of LOCC inequalities~\cite{Plenio2023b}.

\section{Tests using Gaussian and non-Gaussian states}

Recognizing entanglement mediated by gravitational interaction as a pivotal test for probing the quantum aspects of gravity, the prospect of observing such a phenomenon within a controlled laboratory setting arises naturally. This ambitious experiment necessitates the quantum manipulation of two sufficiently massive bodies---an endeavour well-suited for the realm of quantum optomechanics. Over the last decade, quantum optomechanics has made remarkable strides, ushering in a new era of preparing sizable objects in quantum states and skilfully manipulating them coherently. Particularly noteworthy is the subfield of levitated optomechanics, where mechanical degrees of freedom belong to objects suspended in a vacuum, promising an unparalleled degree of environmental isolation by avoiding direct contact with surroundings and enabling in principle unlimited delocalisation~\cite{Stickler2020, Romero-Isart2021a}. In view of these promising opportunities, several targeted proposals have emerged, outlining explicit strategies for the detection of gravitationally induced entanglement using levitated particles. These proposals can be broadly categorised into two groups based on whether they involve the preparation of the massive objects' centre of mass in Gaussian states or not.
\begin{figure}[htbp]
\begin{center}
\includegraphics[width=0.91\textwidth]{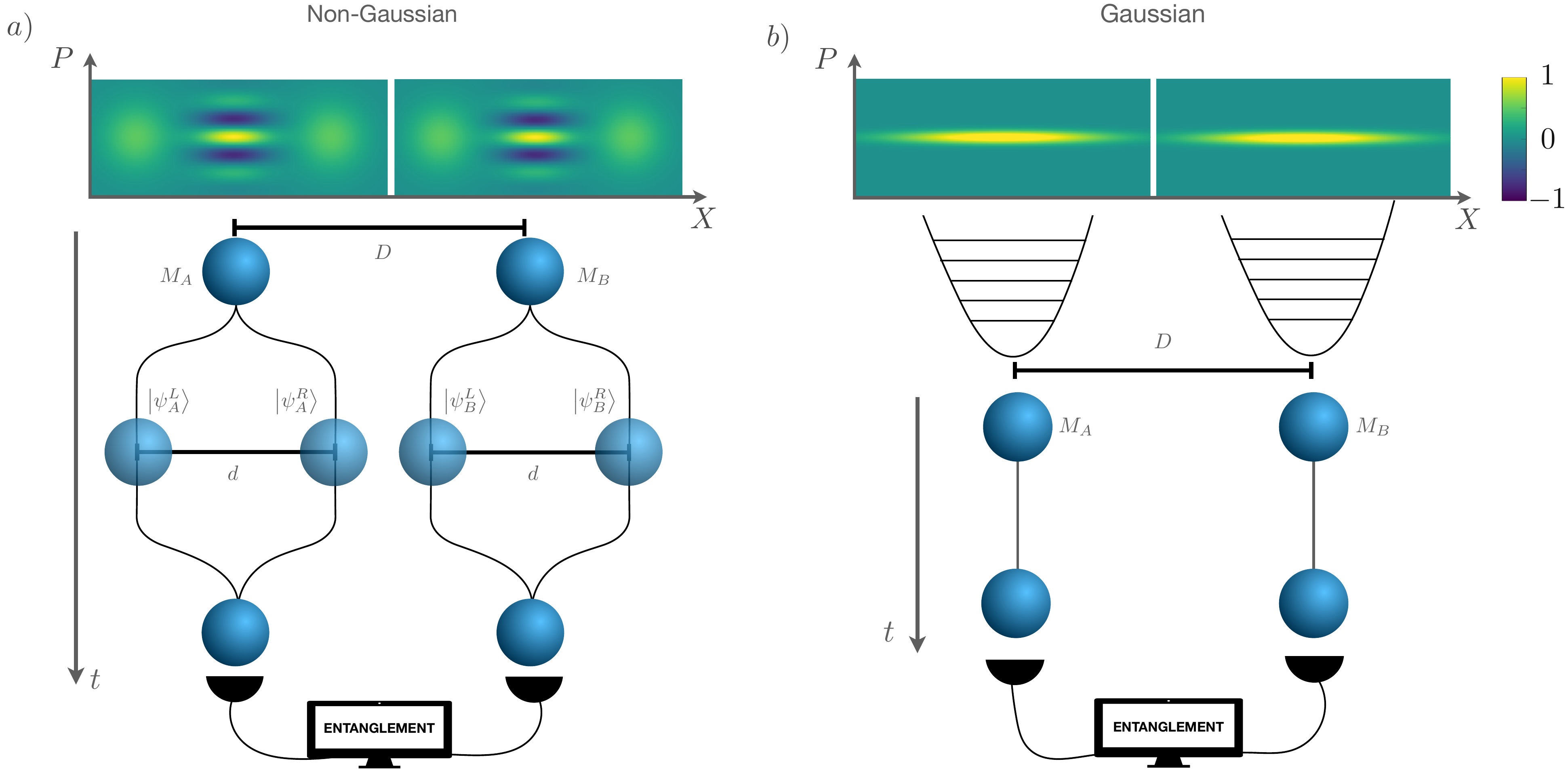}
\caption{{\bf Non-Gaussian and Gaussian tests.} a) Sketch of a test using non-Gaussian states, where the two interacting bodies are prepared in a superposition of two spatially different locations.  The gravitational interaction induces different phases in the different components in superposition leading to the generation of entanglement. b) Sketch of a Gaussian test, where the interacting masses remain always in Gaussian states that are widely delocalised in position---here, the particles are depicted as harmonic oscillators, but can also be free particles, as the free-energy Hamiltonian ${\hat p}^2/(2m)$ preserves their {\it Gaussianity}. Similarly to a), the gravitational interaction generates entanglement between the position degrees of freedom of the two particles, which can then be retrieved by making correlated position and momentum measurements of the two particles. In both drawings, the upper panels show the phase-space representation of each particle. }
\label{fig:phaseVSmomentum}
\end{center}
\end{figure}

\subsection{Tests with non-Gaussian states}
\label{sec:nonGaussian}

Consider two bodies with masses $M_A$ and $M_B$, levitated such that their centres of mass are located a distance $D$ away from each other along some direction $x$. We now prepare the centre of mass of each object in a coherent quantum superposition of two locations separated by a distance $d$ along the same line that connects the two bodies, see Fig.~(\ref{fig:phaseVSmomentum}a). Each body is in a state sometimes referred to as a Schr\"odinger-cat-like state, for it represents a superposition of two macroscopically distinct states. This is a manifestly non-Gaussian state, meaning that its representation in phase space, given by the Wigner function, cannot be expressed in the form of a Gaussian function. In practice, such a state could be prepared by use of an ancillary system, such as a spin that couples to the centre-of-mass motion of the object, e.g. through the presence of a magnetic-field gradient, or some type of non-linearity introduced by a potential field, e.g. a double-well potential.  The exact way in which such a state is prepared is not relevant to the argument, and thus, we will just assume that such a state can be prepared. Under this assumption, the state $|\psi(t=0)\rangle$ of the combined, two-body system at time $t=0$ can be expressed by
\begin{equation}
\label{InterferometricInitial}
    |\psi(t=0)\rangle = \frac{1}{2} (\ket{\psi_A^L} + \ket{\psi_A^R}) \otimes (\ket{\psi_B^L} + \ket{\psi_B^R}),
\end{equation}
where $\ket{\psi_{A(B)}^L}$ and $\ket{\psi_{A(B)}^R}$ describe the state of particle $A(B)$ being in the left or right location, respectively.

We now assume that the particles can only interact through their mutual gravitational interaction, and thus, that any other form of interaction, such as, for example, interaction through Casimir forces, has been reduced sufficiently to be negligible ~\cite{Pedernales2019}. We are interested in finding out how fast entanglement mediated by gravity would build up between two particles in such a configuration, provided, of course, that gravity can indeed mediate entanglement. One way to estimate this, in the low-energy limit considered here, is to add to the system Hamiltonian a Newtonian potential where position variables are replaced by operators, that is,
\begin{equation}
    \hat V(\hat x_A, \hat x_B) = - \frac{G M_A M_B}{\abs{ D + \hat x_B - \hat x_A}},
\label{eq:Newton}
\end{equation}
where $\hat x_A$ and $\hat x_B$ describe the position of each particle as a displacement from their equilibrium positions, see Fig.~(\ref{fig:phaseVSmomentum}). Indeed, in this low-energy regime, the quantisation of the gravitational field is possible and standard techniques agree with a description of this sort~\cite{Taylor2019}. Ignoring the acceleration of the particles due to the gravitational interaction, typically, the transformation of the state in Eq.~(\ref{InterferometricInitial}) under gravity is described by estimating the relative phases that build up for each component of the wavefunction due to the action of gravity, that is,
\begin{equation}
\label{Interferometric}
    \ket{\Psi (t)} = \frac{1}{2} (e^{i \phi_{LL}} \ket{\psi_A^L}\otimes \ket{\psi_B^L} + e^{i \phi_{LR}} \ket{\psi_A^L}\otimes \ket{\psi_B^R} + e^{i\phi_{RL}}\ket{\psi_A^R} \otimes \ket{\psi_B^L} +e^{i \phi_{RR}}\ket{\psi_A^R} \otimes \ket{\psi_B^R}),
\end{equation}
where 
\begin{equation}
    \phi = \phi_{LL} = \phi_{RR} = \frac{GM_AM_B}{\hbar D}t, \qquad \phi_{RL} = \frac{GM_AM_B}{\hbar (D - d)}t \qquad \text{ and } \qquad \phi_{LR}= \frac{GM_AM_B}{\hbar (D + d)}t.
\end{equation}
Factorizing a global phase $\phi$, the state in Eq.~(\ref{Interferometric}) can be rewritten as
\begin{equation}
\ket{\Psi (t)} = \frac{1}{\sqrt{2}}e^{i \phi} \left[\ket{\psi_A^L}\otimes \frac{\ket{\psi_B^L} + e^{i( \phi_{LR} - \phi)}  \ket{\psi_B^R}}{\sqrt{2}} + \ket{\psi_A^R} \otimes \frac{e^{i(\phi_{RL} - \phi)} \ket{\psi_B^L} + \ket{\psi_B^R}}{\sqrt{2}}\right].
\end{equation}
It becomes evident that as soon as $\Delta \phi = \phi_{RL} + \phi_{LR} - 2 \phi \neq 2\pi n$, where $n$ is an integer, the system is not in a product state and hence it is entangled. Certainly, in a more quantitative manner, we can assess the degree of entanglement by calculating an appropriate entanglement measure. In this context, we opt for the logarithmic negativity~\cite{Plenio2005}, which is defined as
\begin{equation}
    E_N(\rho) = \log_2 \norm{\rho^{\Gamma_B}}_1,
\label{eq:logneg}
\end{equation}
where $\norm{\cdot}_1$ is the trace norm and $\rho^{\Gamma_B}$ represents the density matrix of the composite system partially 
transposed w.r.t. to subsystem B, with matrix elements $\rho^{\Gamma_B}_{ij,kl} = \rho_{il,kj}$. To compute $E_N (\rho)$, we treat the problem in the finite subspace of dimension 4 spanned by $\{ \ket{\psi_A^L}, \ket{\psi_A^R} \} \otimes  \{ \ket{\psi_B^L}, \ket{\psi_B^R} \}$. Under the simplifying assumption that $\braket{ \psi_{A(B)}^L | \psi_{A(B)}^R} \approx 0$, a straightforward computation yields
\begin{equation}
    E_N(\rho) = \log_2 (1 + \abs*{\sin \frac{\Delta \phi}{2}} ).
\end{equation}
Of particular interest is the regime $d \ll D$ that is to be expected in experimental implementations, as it guarantees that the surface-to-surface distance between the interacting objects is always large enough to neglect surface forces, such as the Casimir-Polder force that would violate the requirement of having only gravitational interactions. In this regime,
\begin{equation}
    \Delta \phi  \approx \frac{2G M_A M_B t}{\hbar D^3} d^2,
\end{equation}
showing that, at short times, entanglement increases linearly in time at a rate that grows with the product of the mass times the position variance of each particle and decreases with one over the cube of the distance between the particles, that is,
\begin{equation}
    E_N (\rho) \approx \frac{4G M_A M_B t}{\hbar D^3 \ln 2} \Delta x^2.
\label{eq:nonGaussianEnt}
\end{equation}
Notice that in this last expression, we are providing the entanglement in terms of the position variance of the interacting particles, which will prove convenient for comparison with the case studied in the next section. For a cat-like state where the superposed states are separated by a distance $d$, the position variance is given by $\Delta x = d/2$.

\subsection{Tests with Gaussian states}

Let us now consider an alternative configuration, where the interacting particles are not prepared in a Schr\"odinger-cat-like state, but rather they remain in a Gaussian state. Gaussian states are completely characterised by their first- and second-order moments, which are compactly captured by the displacement vector and the covariance matrix,
\begin{equation}
D_i = \expval*{\hat{r_i}} \qquad   \text{ and } \qquad \sigma_{ij} = \expval*{\hat{r_i} \hat r_j +  \hat r_j \hat r_i} - 2\expval*{\hat r_i}\expval*{\hat r_j},
\label{eq:disCov}
\end{equation}
where $\hat r_i$ are the elements of the operator-valued vector $\hat{\bold r} = (\hat X_A,  \hat P_A, \hat X_B, \hat P_B)^T$ containing the dimensionless position and momentum operators of both particles with canonical commutation relations $[\hat X, \hat P] = i$. Such dimensionless operators relate to their dimensional counterparts through $\hat x = x_0 \hat X$ and $\hat p = ( \hbar/x_0) \hat P$, where $x_0$ is an arbitrary constant with dimension of length. 
This allows us to write a dimensionless covariance matrix where all matrix elements have the same units. For example, two initially uncorrelated particles, with the same dimensionless position and momentum variance $\Delta X$ and $\Delta P$, and no correlation between their momentum and position, would have a diagonal covariance matrix
\begin{equation}
\sigma = \left( \begin{array}{cccc} 2 \Delta  X^2 & 0 & 0 & 0 \\ 0 & 2 \Delta P^2 & 0 & 0 \\ 0 & 0 & 2 \Delta X^2 & 0 \\ 0 & 0 & 0 & 2 \Delta P^2 \end{array} \right).
\label{eq:covariance}
\end{equation}
In the quantum optics jargon, this would represent, for each particle, a thermal state of its vibrational mode that can be squeezed in either the position or the momentum quadrature if ${\Delta X^2<1/2}$ or ${\Delta P^2<1/2}$, respectively\footnote{Notice that squeezing of a quadrature different from $\hat X$ or $\hat P$ would generate correlations between momentum and position and thus would not lead to a diagonal covariance matrix}. We want to see if such a configuration can reach an entanglement rate comparable to that of the non-Gaussian states discussed earlier. The particles interact through the same Newtonian potential term in Eq.~(\ref{eq:Newton}), which expanded into powers of the just introduced dimensionless operators, reads 
\begin{equation}
V(\hat X_A, \hat X_B) = - G\frac{M_AM_B}{D} + G \frac{M_AM_Bx_0}{D^2}(\hat X_A - \hat X_B) - G \frac{M_A M_B x_0^2}{D^3}(\hat X_B - \hat X_A)^2 + ...\, .
\end{equation}
Provided that the variance in position of each particle satisfies $\Delta X_A, \Delta X_B \ll D/x_0$, the expansion can be truncated to a few lower orders. Notice that this is the same regime considered in our earlier example to arrive at the expression in Eq.~(\ref{eq:nonGaussianEnt}). In the expansion, we are interested in identifying the first term that can, at least in principle, generate entanglement between the two particles. This should be a non-local term, in the sense that it should contain a product of operators acting in the Hilbert spaces of particle A and particle B. In our case, this is a second-order term with the form $2 \frac{G M_A M_B x_0^2}{D^3} \hat X_A \hat X_B$. We thus truncate the expansion at second order and consider the following exercise: initialise the system in a state described by Eq.~(\ref{eq:covariance}), evolve it under a Hamiltonian $\hat H =  \lambda \hat X_A \hat X_B$ where $\lambda \equiv 2GM_AM_B x_0^2/D^3$, and monitor the entanglement build up between particles $A$ and $B$. This is of course a bit different from what a practical realisation would look like, as in practice the Hamiltonian would contain terms other than the entanglement-generating one. For example, if the particles are trapped in a harmonic potential, one would need to include a free-energy Hamiltonian $\hat H_0 = \hat p^2 / (2m) + \frac{1}{2}m \omega^2 \hat x^2$ for each particle, which would modify the dynamics, including the evolution of entanglement. However, this is not different from the approach we have taken in the previous example where the particles were held in a particular state and only the effect of their interaction was analysed while ignoring other local dynamics. We, here, want to isolate the effect of gravitational interaction and understand its ability to generate entanglement depending on the initial state, while the effects of other local dynamics that are particular to the chosen implementation, shall be discussed elsewhere. 

A Hamiltonian of the form $\hat H =  \lambda \hat X_A \hat X_B$ is quadratic in position and momentum operators and thus it follows that an initially Gaussian state evolving under such a Hamiltonian remains Gaussian at all times. This allows us to describe the system by its covariance matrix at all times~\cite{Serafini}. In fact, the evolution of the covariance matrix can be easily solved by computing the time evolution of the vector $\hat {\bold r}(t) = (\hat X_A(t), \hat P_A(t), \hat X_B(t), \hat P_B(t))^T$ in the Heisenberg picture, and then constructing the covariance matrix according to Eq.~(\ref{eq:disCov}). In the particular case at hand, the solution is given by 
\begin{equation}
\sigma (t) = e^{Kt} \sigma (0) e^{K^T t} \qquad \text{with} \qquad K = \left( \begin{array}{cccc} 0 & 0 & 0 & 0\\ 0 & 0 & -\lambda/\hbar  &0 \\ 0 & 0 & 0 & 0 \\ -\lambda/\hbar & 0 & 0 & 0 \end{array}  \right).
\end{equation}
For an initial state such as that in Eq.~(\ref{eq:covariance}), this results in a covariance matrix at time $t$ 
\begin{equation}
\sigma (t) = \left(
\begin{array}{cccc}
 2\Delta X^2 & 0 & 0 & - 2\Delta X^2 \lambda  t/\hbar \\
 0 & 2 \Delta P^2+ 2 \Delta X^2 \lambda ^2 t^2/\hbar^2 & - 2 \Delta X^2 \lambda  t /\hbar & 0 \\
 0 & - 2 \Delta X^2 \lambda t / \hbar & 2 \Delta X^2 & 0 \\
 - 2 \Delta X^2 \lambda t / \hbar & 0 & 0 & 2 \Delta P^2+ 2 \Delta X^2 \lambda^2 t^2 / \hbar^2 \\
\end{array}
\right).
\end{equation} 
We are now left with the task of computing the entanglement of $\sigma (t)$. To that end, we want to look again at the logarithmic negativity introduced in Eq.~(\ref{eq:logneg}), which for this two-mode Gaussian system, can be computed in terms of its covariance matrix (see Eq.~(7.47) in \cite{Serafini}) via
\begin{equation}
    E_N(\sigma) = \max [0, -  \log_2(\nu_{\rm min})],
\end{equation}
where $\nu_{\rm min}$ is the smallest symplectic eigenvalue of the covariance matrix $\tilde \sigma$ of the partially transposed density matrix associated with $\sigma$. The symplectic eigenvalues of $\tilde \sigma$ are obtained as the absolute values of the standard eigenvalues of matrix $i\Omega \tilde \sigma$, where $\Omega$ is the symplectic matrix, whose elements are given by the commutation relations $\Omega_{ij} =  -i [\hat r_i, \hat r_j]$. A straightforward way to find the matrix $\tilde \sigma$ is to invert the sign of the momentum of one of the particles in matrix $\sigma$. For instance, for partial transposition w.r.t. particle $B$, the desired transformation is achieved by $\tilde \sigma = P \sigma P$, with $P = \text{diag}{(1,1,1,-1)}$. Putting everything together, we find 
\begin{equation}
    \nu_{\rm min}^2 = 4 \Delta X^2 \Delta P^2 + 8 \eta(t) [\eta(t) - \sqrt{\Delta X^2 \Delta P^2 + \eta(t)^2}], 
\end{equation}
where $\eta(t) \equiv \Delta X^2 \lambda t /\hbar$. If we consider a pure initial state, that is a squeezed vacuum state, for which $\Delta X \Delta P = 1/2$, and for times  such that $ \eta(t) \ll 1/2$, the expression can be simplified to
\begin{equation}
\nu_{\min}^2 = 1 - 4 \eta(t) + O(\eta(t)^2).
\end{equation}
The system is entangled as long as $\nu_{\rm min} < 1$, which occurs for all positive times $t>0$ because $\eta(t)$ is always positive. Thus, at short times the logarithmic negativity of the gravitationally-interacting two-particle system in Gaussian states evolves as  
\begin{equation}
    E_N(t) = - \frac{1}{2} \log_2(1 - 4 \eta (t)) + O(\eta(t)^2) = \frac{2 \eta(t)}{\ln 2} + O(\eta(t)^2).
\end{equation}
We see that the rate of change of the entanglement is again given by 
\begin{displaymath}
    \frac{dE_N(t)}{dt} = \frac{2 \lambda \Delta X^2}{\hbar\ln 2} = \frac{4 G M_A M_B }{\hbar D^3\ln 2}\Delta x^2,   
\end{displaymath} 
which is the same rate we found for the case of non-Gaussian states in Eq.~(\ref{eq:nonGaussianEnt}). 

\section{The phase-space picture}
In the introduction of this article, we announced that we would study and compare the force sensitivity of phase-based with momentum-based force detectors. However, so far we have only talked about the entanglement rate between gravitationally interacting bodies, and thus, the reader might be wondering what the relationship between entangling rate and force sensitivity is. In fact, these two are very closely related. Consider the case of two particles, where particle $A$ is prepared in a Schr\"odinger-cat-like state, as the ones discussed earlier, $\ket{\psi_A^L} + \ket{\psi_A^R}$, while particle $B$ remains in an arbitrary state $\ket{\psi_B}$. The evolution of the combined state under their mutual gravitational interaction can be studied separately for each of the components of the state of particle $A$, that is,
\begin{equation}
\begin{array}{cc} 
    \ket{\psi_A^L} \otimes \ket{\psi_B} \rightarrow \ket{{\psi_A^L}'} \otimes \ket{\psi_B'}\, , \\[0.5ex]
    \ket{\psi_A^R} \otimes \ket{\psi_B} \rightarrow \ket{{\psi_B^R}'} \otimes \ket{\psi_B''}.
\end{array}
\end{equation}
Here, $\ket{\psi_B'}$ represents the state that particle $B$ assumes when subjected to the gravitational influence of those components of the wave-function of particle $A$ that are localised around the left position, while $\ket{\psi_B''}$ represents the state arising from the gravitational interaction with the components that are localised around the right position. As already discussed, when $\braket{\psi_B' | \psi_B''} < 1$, the particles are entangled. On the other hand, the sensitivity of particle B to a force, can be understood as its ability to evolve under the influence of such a force into a state that is distinguishable from the original one. With this in mind, the relation between entanglement and force sensitivity becomes clear: the system's entanglement sensitivity characterises how particle $B$ can discern variations in the gravitational forces originating from the distinct positions of particle $A$. It is, thus, instructive to learn about the force sensitivity of states to infer their ability to get entangled. To that end we would like now to consider the force sensitivity of particle $B$ depending on whether it has been prepared initially in a cat state or a Gaussian squeezed state. This allows us to make a connection with the two scenarios discussed earlier for gravitationally generated entanglement.

\begin{figure}[htbp]
\begin{center}
\includegraphics[width=1\textwidth]{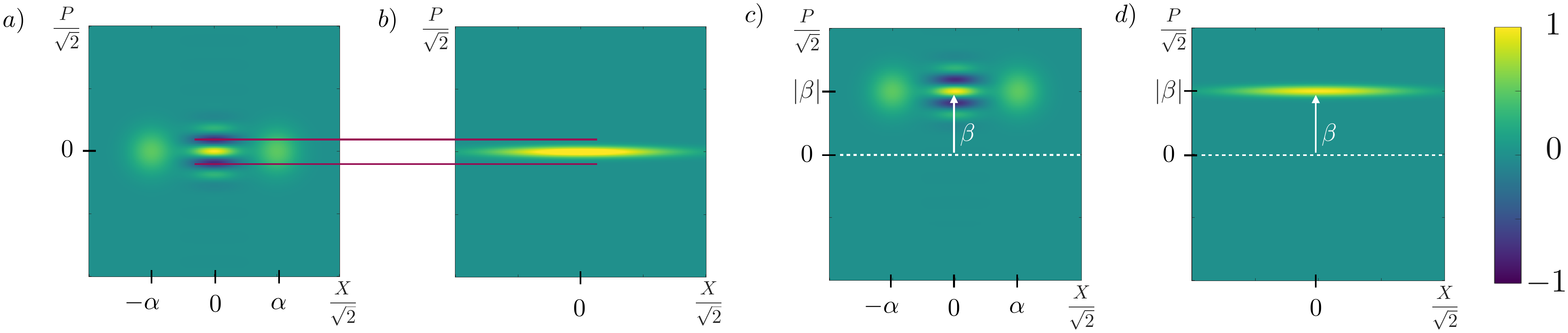}
\caption{{\bf Wigner functions.} a) shows the Wigner function of a Schr\"odinger-cat state of the form $\propto \ket{\alpha} + \ket{-\alpha}$ and b) a squeezed state with the same spatial delocalization $\Delta X = \sqrt{2} \alpha$. The red lines indicate the relation between the periodicity of the interference pattern of the cat state and the variance in momentum of the squeezed state.  c) and d) show the Wigner functions of the same states after transformation under a purely imaginary displacement operation.}
\label{fig:Wigner}
\end{center}
\end{figure}

We study the force sensitivity of these two states via their phase-space representation, more specifically their Wigner function.
The Wigner function is a quasiprobability distribution that offers a complete representation of the quantum state of a continuous variable system \cite{GerryK2023}. That is, it contains as much information as the density matrix of the same system, and thus, allows to compute the expectation value of any observable. It has the added appeal that it can be drawn in a 2-dimensional plot whose coordinates represent position and momentum and thus offers a visual aid that will allow us to gain some intuition on the physics involved in determining the force sensitivity of quantum states. For a general quantum state $\rho$, the Wigner function is defined as
\begin{equation}
    W(X,P) = \frac{2}{\pi} \int_{\mathbb{R}} dy\, e^{-2iPy} \langle X+y|\rho|X-y\rangle,
\end{equation}
which is given as a function of the dimensionless position and momentum coordinates $X$ and $P$ that are the eigenvalues of the dimensionless position and momentum operators $\hat X$ and $\hat P$. It is worth noting that the marginals of the Wigner function yield the probability distributions
in position and momentum, that is
\begin{equation}
    \langle X|\rho|X\rangle = \frac{1}{2}\int_{-\infty}^{\infty} dP\, W(X,P) \;\;\;\;  \mbox{and} \;\;\;\;
    \langle P|\rho|P\rangle = \frac{1}{2}\int_{-\infty}^{\infty} dX\, W(X,P)\, .
\end{equation}
For the case of a cat state composed of the superposition of two coherent states $|\alpha\rangle = e^{-|\alpha|^2/2}\sum_{n=0}^{\infty} \frac{\alpha^n}{\sqrt{n!}}|n\rangle$, that is $\ket{\psi_{\rm cat}} = C (\ket{\alpha} + e^{i \phi} \ket{- \alpha})$, where $C= (2 + 2 \cos{\phi} e^{-2 \abs{\alpha}^2})^{-1/2}$ is a normalisation constant and $\alpha$ is a complex number, the Wigner function is given by
\begin{equation}
W_{\rm cat} (X,P) = \frac{1}{\pi(1+e^{-2|\alpha|^2})}\left[ e^{-({\bf r}-{\bf r'})^T({\bf r}-{\bf r'})} + e^{-({\bf r}+{\bf r'})^T({\bf r}+{\bf r'})} + 2 e^{-{\bf r}^T{\bf r}} \cos(2{\bf r}^T\Omega{\bf r'} + \phi) \right] ,\\
\label{wignerCat}
\end{equation}
with ${\bf r} = (X,P)^T$, ${\bf r}' = \sqrt{2} (\Re(\alpha),\Im(\alpha) )^T$ and $\Omega = (0\, , 1;-1\, , 0)$ \cite{Serafini}. It contains three important elements: two displaced Gaussians that each corresponds to one of the coherent states in superposition 
and a third term centred about the origin which describes an oscillatory feature in phase space whose period, importantly for the following,  scales inversely proportional to the distance between the superposed coherent states. In the remainder of this section, we will assume that $\alpha$ is real 
which corresponds to a cat that is initially delocalised in position along the x-axis, for an illustration see  Fig.~(\ref{fig:Wigner}a). 

The second state that we are interested in is a Gaussian squeezed state that spreads out widely in space while being very narrow in momentum. We can characterise it conveniently as a squeezed vacuum state denoted as $\ket{\xi} = \hat S(\xi) \ket{0}$. Here, ${\hat S(\xi) = \exp{(\xi^* \hat a^2 - \xi \hat a^{\dag 2})/2}}$ is the squeezing operator with complex squeezing parameter $\xi = r e^{i \theta}$, $\hat a$ and $\hat a^\dag$ are ladder operators defined as $\hat a = (\hat X + i \hat P)/\sqrt{2}$ and its conjugate, and $\ket{0}$ is an eigenstate of $\hat a$ with eigenvalue 0. The Wigner function of such a state reads
\begin{equation}
    W_{\rm sqz} (X,P) = \frac{2}{\pi} e^{-\bf r^T \sigma_{\rm sqz}^{-1} \bf r},
\end{equation}
which, as it is the result of an evolution with a Hamiltonian that is quadratic in position and momentum, is a two-dimensional Gaussian function. Here, $\sigma_{\rm sqz}$ is the covariance matrix of the squeezed vacuum state 
\begin{equation}
    \sigma_{\rm sqz} = \left( \begin{array}{cc} \cosh(2r) - \sinh (2r) \cos (\theta) & -\sinh (2r) \sin (\theta) \\  -\sinh (2r) \sin (\theta) & \cosh(2r) + \sinh (2r) \cos (\theta)  \end{array} \right),
\end{equation}
which has $\det (\sigma_{\rm sqz}) = 1$ and thus its inverse always exists. In the following, we will consider states with covariance matrix $\sigma_{\rm sqz} = {\rm diag}(e^{2r}, \, e^{-2r})$, corresponding to states that are squeezed along their momentum quadrature and thus have an increased delocalisation in position, that is, they have a squeezing parameter with $\theta = \pi$, see Fig.~(\ref{fig:Wigner}b).

Let us now consider the dynamics of each of these two possible states of particle $B$ under the gravitational force exerted by particle $A$. We will again consider the linearised regime discussed before, where the gravitational interaction takes the form $\hat H= \lambda \hat X_A \hat X_B$, and thus leads to a unitary, time-evolution operator $\hat U = \exp{- i \frac{\lambda t}{\hbar} \hat X_A \hat X_B}$. By writing the position operator of particle $B$ in terms of its ladder operators $\hat b$ and $\hat b^\dag$ according to the relation introduced above, we identify that, for each position eigenstate of particle $A$, $\ket{X_A}$, the gravitational interaction results in a displacement of particle $B$
\begin{equation}
    \hat U \ket{X_A} \otimes \ket{\psi_B} = \ket{X_A} \otimes \hat D[\beta(X_a)] \ket{\psi_B}.
\end{equation}
Here, $\hat D(\gamma) = \exp{\gamma \hat a - \gamma^* \hat a^\dag}$ is the displacement operator and we have defined $\beta \equiv -i \lambda X_A t / \ (\hbar \sqrt{2})$ as a purely imaginary displacement parameter that is a function of the position of particle $A$, $X_A$. The effect of the displacement operator on any state is to displace the state in phase space in the direction, and by a magnitude, specified by the complex displacement parameter $\gamma = \gamma_{\rm r} + i \gamma_{\rm im}$, such that its Wigner function transforms as ${W(X,P) \rightarrow W(X - \sqrt{2} \gamma_{\rm r}, P - \sqrt{2} \gamma_{\rm im})}$. In Fig.~(\ref{fig:Wigner}c-d) an imaginary displacement, that is a displacement in the direction of the coordinate $P$, is plotted for the two states of interest. In this context, we will define the force sensitivity of particle $B$ as the rate at which the overlap between its initial state $\ket{\psi_B}$ and its state after a displacement in the coordinate $P$, $\hat D (\beta) \ket{\psi_B}$, decreases.

\begin{figure}[htbp]
\begin{center}
\includegraphics[width=0.91\textwidth]{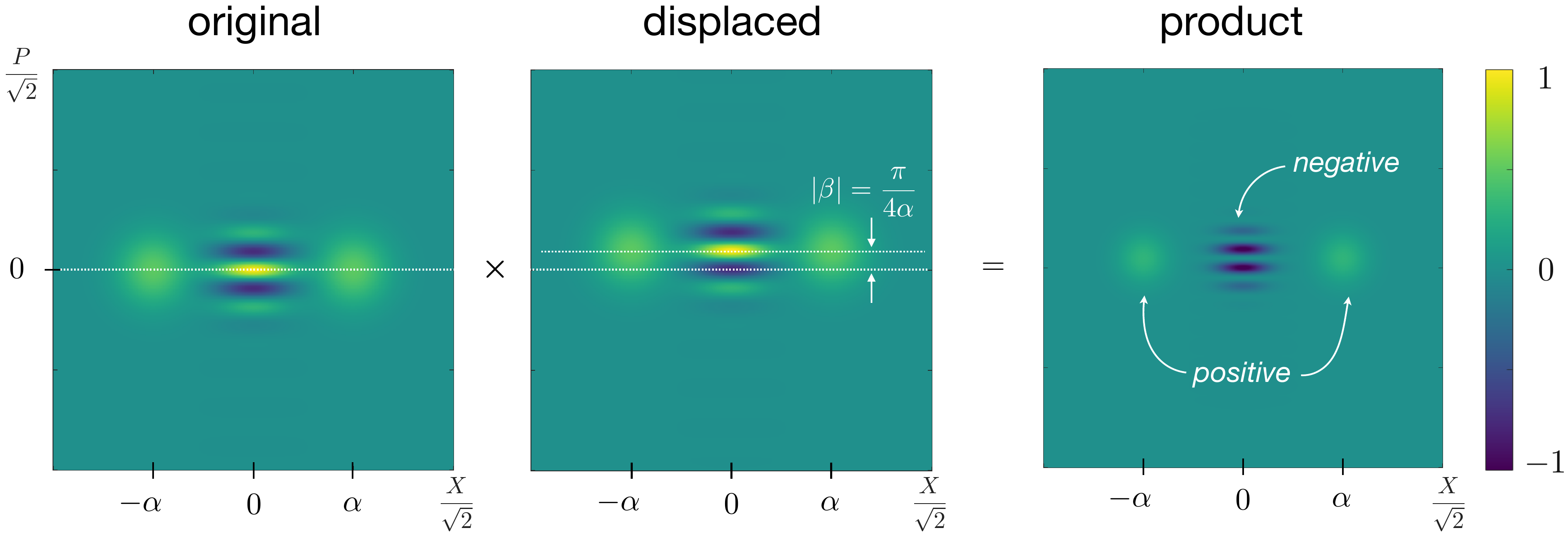}
\caption{{\bf Orthogonality of cat states.}  The left-hand-side plot shows the Wigner function of a cat state of the form $\propto \ket{\alpha} + \ket{-\alpha}$. The plot in the middle shows the Wigner function of the same state after being displaced by $\beta = i \pi/(4 \alpha)$, corresponding to the displacement that shifts the interference pattern by a phase $\pi$. The right-hand-side plot shows the product of the two previous functions. The integral of the product over the entire plane is $0$ indicating that the two states are orthogonal. }
\label{fig:WignerOrthogonality}
\end{center}
\end{figure}

The overlap between any two quantum states $\ket{\psi}$ and $\ket{\phi}$ can be written in terms of their Wigner functions as
\begin{equation} \label{overlap}
	|\langle\psi|\phi\rangle|^2 = \frac{\pi}{2} \int_{\mathbb{R}}   dX dP \, W_{|\psi\rangle}(X,P)\,  W_{|\phi\rangle}(X,P).
\end{equation}
That is, given two Wigner functions, one can compute their overlap by taking a point-by-point product of the two functions and then integrating the result over the entire phase-space plane. In general, the phase-space representation provides a visual way to understand the similarity between two quantum states by examining the overlap between their corresponding Wigner functions. For example, given two squeezed states $\ket{\xi}$ and $\ket{\xi'}$ that differ from each other only by a displacement in phase space, their overlap $\abs{\braket{\xi | \xi'}}^2$ can be visualised as the overlap between two Gaussian functions in phase space. This overlap naturally decreases exponentially as the distance between the peak positions of these Gaussian functions increases. One might be tempted to think that for any given state, the smaller the variance of the Wigner function in the direction of the displacement, the faster the overlap between the original and the displaced state will decrease. Under such a naive logic, one would be led to conclude that the cat state is less sensitive to displacements in momentum than a Gaussian squeezed state, after all, its variance in momentum is much larger than that of a squeezed state with a comparable position delocalisation. However, while such a conclusion would be correct for a statistical mixture of the two coherent states that make up the Schr{\"o}dinger cat, it is wrong here as it would ignore the quantum interference between the two components of the cat state. 

Consider a small displacement in momentum that shifts the Wigner function of a cat state by an amount that moves the interference fringes in the middle region of its phase-space representation by $\pi$ out of phase with those of the original state, that is, a displacement after which the maxima take the position of the minima and vice-versa. If one looks at the overlap between the original and displaced Wigner functions, three regions can be identified: the Gaussians on the sides of the state have nearly perfect overlap with their original counterparts and give a positive contribution in Eq.~(\ref{overlap}) while the interference term gives an equal but negative contribution, see Fig.~(\ref{fig:WignerOrthogonality}). Therefore, the overlap vanishes as the positive contribution from the Gaussians on the sides cancels exactly with the negative contribution from the interference fringes. The sensitivity to displacements of the cat state is, thus, dominated by the distance between its interference fringes and not by the size of the coherent states that compose it. Importantly, as stated earlier, the distance between the fringes reduces in a manner that is inversely proportional to the delocalisation of the cat state, as captured by the third term in Eq.~(\ref{wignerCat}) which is proportional to $\cos{(2 \Delta X_{\rm cat} P)}$. 

Similarly, the squeezed state has a momentum variance that also scales with the inverse of the position variance, as captured by the Heisenberg uncertainty relation $\Delta X \Delta P = 1/2$. A comparison between the two is drawn in Fig.~(\ref{fig:Wigner}a-b). This is an important realisation that becomes evident when looking at the problem in phase space and which clarifies why the sensitivity of cat and squeezed states with the same delocalisation in position is comparable. For a more quantitative account of this, one can compute the overlap of each state with its displaced counterpart
\begin{align}
   \label{eq:catOverlap} \abs{\braket{\psi_{\rm cat} | \hat D(\beta) | \psi_{\rm cat}}}^2 &=  e^{-\abs{\beta}^2} \cos^2{(\sqrt{2} \abs{\beta} \Delta X_{\rm cat})}
\end{align}
and
\begin{align}
   \abs{\braket{\psi_{\rm sqz} | \hat D(\beta) | \psi_{\rm sqz}}}^2 &=  \exp{-  (\sqrt{2} \abs{\beta}\Delta X_{\rm sqz} )^2}.
\end{align}
Here, for simplicity, we have assumed, for the cat state,  that $\abs{\braket{\alpha | - \alpha}}^2 \approx 0$ and $\Delta X_{\rm cat} \approx \sqrt{2} \alpha$, which are satisfied provided that $\abs{\alpha}^2 >> 1/4$. 
Naturally, the two do not behave exactly in the same manner. While the overlap between the cat states follows a squared cosine function, the overlap between squeezed states follows an exponential decay. Nevertheless, for small $\abs{\beta}$ they both scale as $1 - (\sqrt{2} \abs{\beta}\Delta X)^2$ and thus depend in the same way on their delocalisation. We, therefore, confirm that the relevant resource in the sensitivity to force is the delocalisation in position $\Delta X$. This is not all that surprising as we had already established that the entanglement rate between two squeezed states and two Schr\"odinger-cat-like states is the same. However, the attentive reader might realise that when computing the entanglement rate of the two cat states, we did not consider the complete evolution of the system with its associated displacements due to the mutual gravitational interaction, but rather just the build-up of relative phases, and yet we had reached the same conclusion, namely, that for equal delocalisation, squeezed Gaussian states and cat states get entangled at the same rate. Thus, we would now like to look at what the dynamical-phase approach yields in the phase-space representation.

\begin{figure}[htbp]
\begin{center}
\includegraphics[width=0.6\textwidth]{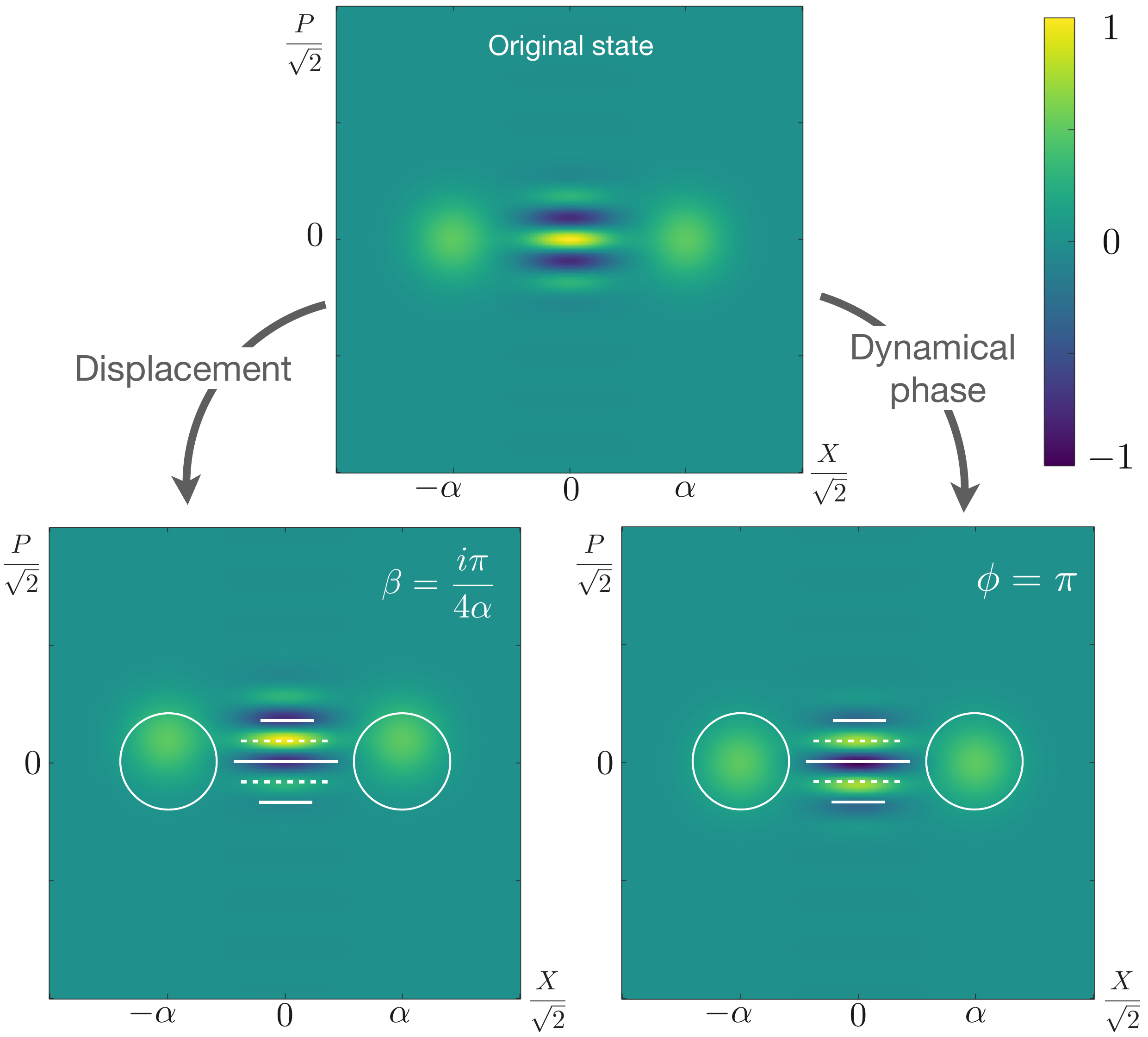}
\caption{{\bf Relative phase VS displacement.} Evolution of the Schr\"odinger-cat state (upper panel) under the action of a displacement operator, as computed by the dynamical-phase approach (bottom-right panel) and by the displacement-in-phase-space approach (bottom-left panel). The white lines in the bottom row panels indicate schematically the phase-space distribution of the original state. Circles represent the position of the two superposed coherent states, while lines indicate the position of the maxima (continuous lines) and minima (dashed lines) of the interference fringes. We observe that in both cases the transformed interference fringes become perfectly out of phase with the original ones, leading to the orthogonality between the original and transformed states. On the other hand, the overlap of the coherent states on the sides with their original counterparts is perfect for the dynamical-phase case, while it drops slightly for the displaced case. Nevertheless, when the displacement is small enough these two approaches yield the same estimation of the overlap.}
\label{fig:PhaseVSDisplacement}
\end{center}
\end{figure}

The dynamical-phase approach employed in Section~(\ref{sec:nonGaussian}), stipulates that the evolution of a Schr\"odinger-cat state can be approximated by multiplying each of the states in superposition with a phase $\exp{- i \expval*{\hat H}t/\hbar}$, where the expectation value of the Hamiltonian is taken over the corresponding component of the wave function. For each position eigenstate of particle $A$, the Hamiltonian experienced by particle $B$ can be expressed as $\hat H = \lambda X_A \hat X_B$. Then, taking into account that $\bra{\pm \alpha} \hat X \ket{\pm \alpha}= \pm \sqrt{2} \alpha$, the cat state would transform as
\begin{equation}
\frac{\ket{\alpha} + \ket{-\alpha}}{\sqrt{2}} \rightarrow \frac{e^{  \sqrt{2}  \beta \Delta X}\ket{\alpha} + e^{- \sqrt{2}  \beta \Delta X} \ket{-\alpha}}{\sqrt{2}},
\label{eq:relPhase}
\end{equation}
where we have expressed the phases in terms of the earlier introduced parameter ${\beta = -i \lambda X_A t /(\hbar \sqrt{2})}$, and the position variance of the cat state $\Delta X = \sqrt{2} \alpha$. The state retains the form of a cat state centred around the origin of the phase space but now acquires a relative phase  $2 \sqrt{2}  \abs{\beta} \Delta X$ between its two components. The effect of such a phase is to shift the interference fringes in the central region of the phase-space representation of the cat state as described by Eq.~(\ref{wignerCat}), leading to a state that is orthogonal to the original one when the phase reaches a value of $\pi$. 

The key aspect here, which becomes evident in this phase-space description of the problem, is that the pace at which the interference pattern shifts in phase space is the same if one considers the approximate approach of dynamical phases or the full action of the Hamiltonian, that is, the displacement of the entire cat state in momentum $P$, see Fig.~\ref{fig:PhaseVSDisplacement}. In fact, the overlap between the left- and right-hand-side states in Eq.~(\ref{eq:relPhase}) can be easily computed and is given by $\cos^2{(\sqrt{2} \abs{\beta} \alpha)}$, which is the same as in Eq.~(\ref{eq:catOverlap}) except for the factor $e^{- \abs{\beta}^2}$. This factor quantifies precisely the contribution from the loss of overlap between the coherent states, which can only be captured if the displacement of the state is considered. However, for displacements much smaller than the coherent state width $\beta \ll 1/\sqrt{2}$, which are generally sufficient to make the state orthogonal to its undisplaced original, the contribution of this factor is negligible, and accounting only for the relative phase build-up provides an accurate computation of the overlap between the original and the transformed states, and therefore, it also serves as a valid estimator of the entanglement between the two particles. Clearly, this occurs because the sensitivity of the cat state originates from the structure of its interference fringes rather than from the coherent states that compose it. However, one should be cautious because following the same procedure could lead to wrong estimates if instead of coherent states in superposition, one would deal with other states in superposition, which individually have a sensitivity to displacements in $P$ that is comparable to that of its interference fringes. This is the case, for example, of squeezed-cat states.  

	\begin{figure}[hbt]
		\begin{center}
			\includegraphics[width=1 \linewidth]{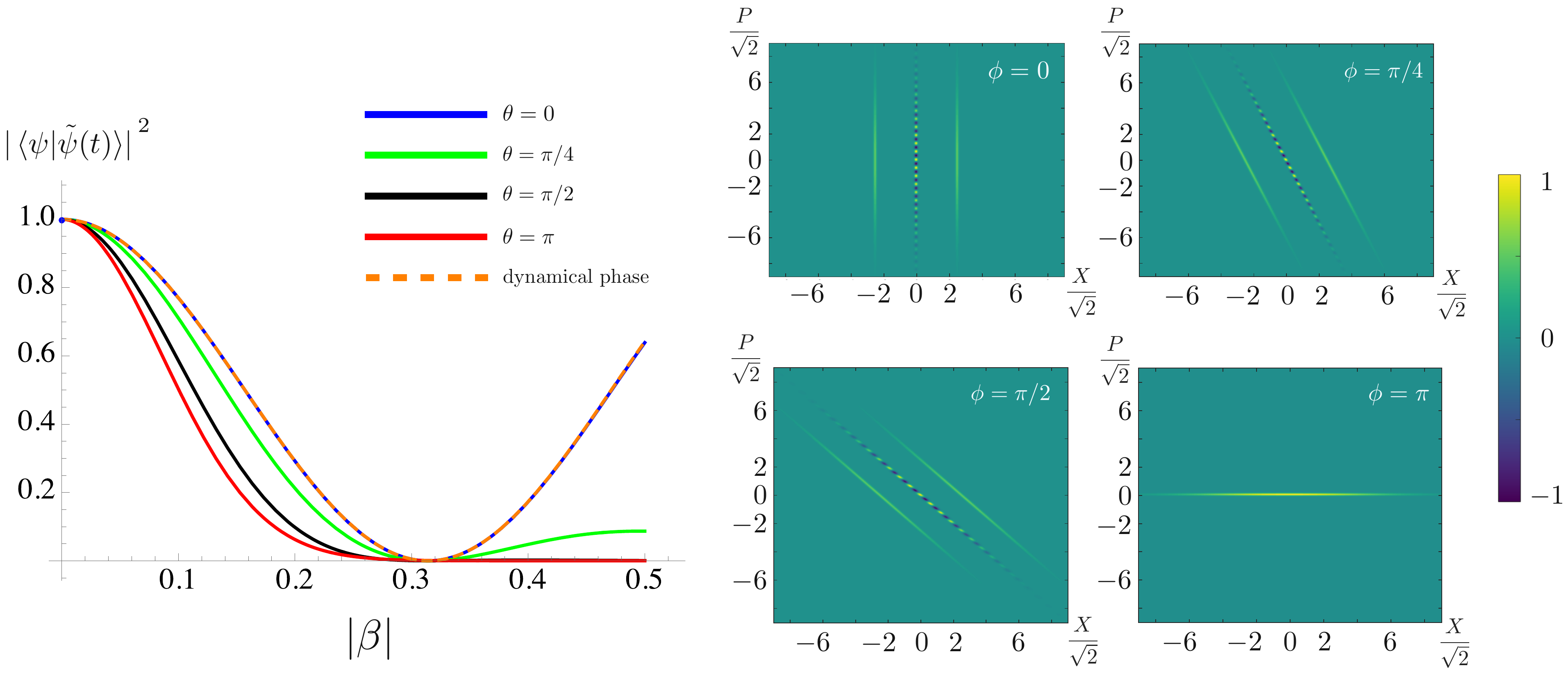}
		\end{center}
		\caption{{\bf Squeezed cat states.} The plot on the left shows the overlap of a squeezed Schr\"odinger-cat state with a displaced copy of itself as a function of the purely imaginary displacement parameter $\beta$. We consider a collection of states with parameters $r=2$ and $\alpha = 2.5$, and several squeezing angles that exhibit varying sensitivity. The corresponding Wigner functions are plotted in the panels to the right. For reference, the overlap computed according to the dynamical-phase approach is also plotted as an orange dashed line. The results show that the dynamical-phase approach underestimates how fast the overlap drops as a function of $\beta$ for all squeezed cat states with $\theta \neq 0$. This is a consequence of the dynamical-phase approach ignoring the displacement in $P$ of the squeezed states that make up the squeezed cat state.}
		\label{Fig2}
	\end{figure}
	
Consider a squeezed vacuum state $\ket{\xi}={\hat S}(\xi)|0\rangle$ that is prepared in a superposition of being displaced by an amount $\alpha$ in two opposite directions along the position quadrature $X$, that is 
\begin{equation}
\label{squeezedCat}
    \ket{\psi} = \frac{1}{N} (\hat D(\alpha) \ket{\xi} + \hat D(- \alpha) \ket{\xi}),
\end{equation}
where $N = \sqrt{2 (1 + \exp\{-2 \alpha^2 |{\cosh(r) + \sinh(r) e^{i \phi}}|^2\}}$ is a normalisation constant, and $\alpha = \alpha^*$. Again, we want to compute how fast the state becomes orthogonal to its initial version under a purely imaginary displacement. If we follow the relative-phase argument described earlier, the transformed state would look like 
\begin{equation}
\label{squeezedCatPhase}
    \ket{\psi(t)} = \frac{1}{N} (e^{ 2 \beta \alpha } \hat D(\alpha) \ket{\xi} + e^{- 2 \beta \alpha}\hat D(- \alpha) \ket{\xi}).
\end{equation}
Alternatively, if we consider the action of the displacement operator, the evolved state takes the form
\begin{equation}
\label{squeezedCatDisplacement}
    \ket{\smash{\tilde \psi (t)}} = \hat D(\beta) \ket{\psi} \\= \frac{1}{N} (e^{\beta \alpha} \hat D(\alpha + \beta) \ket{\xi} + e^{-\beta \alpha } \hat D(-\alpha +  \beta ) \ket{\xi} ),
\end{equation}
where we have used the composition rule for displacement operators ${\hat D(\beta) \hat D(\alpha) = \exp\{(\beta \alpha^* - \beta^* \alpha)/2 \} \hat D(\beta + \alpha)}$. The overlap between $\ket \psi$ and the state $\ket{\psi (t)}$ computed through the dynamical-phase approach in Eq.~(\ref{squeezedCat}) is given by 
\begin{equation}
    \abs{\braket{\psi | \psi (t)}}^2 = \cos^2(2 \abs{\beta} \alpha),
\end{equation}
which is the same as that computed for the {\it non-squeezed} cat state through the same approach. This is not surprising because this approach only considers the distance between the interference fringes, which is the same for squeezed and standard cat states. On the other hand, the overlap between $\ket{\psi}$ and $\ket{\smash{ \tilde \psi (t)}}$ in Eq.~(\ref{squeezedCatPhase}) is then given by 
\begin{align}
\label{overlapDisp}
    \abs{\braket{\psi|\smash {\tilde \psi (t)}}}^2 &= \frac{4 e^{- \abs{\beta}^2 \abs{\cosh(r) - \sinh(r) e^{i \theta}}^2}}{N^4} \{  \cos(2 \abs{\beta} \alpha ) + e^{ - 2 \alpha^2 \abs{\cosh(r) + \sinh(r) e^{i \theta}}^2} \cos [ 2 \abs{\beta} \alpha \sinh(2r) \sin(\theta)]  \}^2 \\ &\approx e^{- \abs{\beta}^2 \abs{\cosh(r) - \sinh(r) e^{i \theta}}^2} \cos^2(2 \abs{\beta} \alpha ),
\end{align}
where the approximation in the second line corresponds to neglecting the overlap between the different components of the cat state, that is, between squeezed states that are displaced in opposite directions, and which is valid provided that $2 \alpha^2 \abs{\cosh(r) + \sinh(r) e^{i \phi}}^2 \gg 1$. It becomes evident that, for a generic squeezed cat state, the two approaches yield rather different results. In particular, for certain squeezing parameters the relative-phase approach would underestimate the sensitivity of the state to small forces, because, in this case, the factor $e^{- \abs{\beta}^2 \abs{\cosh(r) - \sinh(r) e^{i \phi}}^2}$ might not be negligible anymore.  Naturally, this would also lead to a wrong estimation of the amount of entanglement in the state. In Fig.~(\ref{Fig2}), we plot Eq.~(\ref{overlapDisp}) for a collection of squeezed cat states with different squeezing angles $\theta$, and confirm that the relative-phase approach underestimates the sensitivity for all states with $\theta \neq 0$ and that the error grows with increasing values of $\theta$. In the phase-space picture, it becomes clear that this occurs because, under a displacement, the overlap between the squeezed states that make up the cat state can drop faster than the overlap between the interference fringes, something that the dynamical-phase approach cannot capture. 

\section{Conclusions}
To unveil the quantum characteristics of gravity in the low-energy domain, it is essential to attain an unparalleled level of force sensitivity using mechanical systems.
However, the sensitivity of mechanical systems depends on how quickly a force can transform the quantum state of the particle into a state that is distinguishable from the original one. Naturally, different states will lead to different sensitivities. We have shown that the action of a homogeneous force equates to a displacement of the state in phase space in the direction of momentum. Under this notion we were able to establish that for Schrödinger-cat and Gaussian squeezed states, sensitivity is determined by the state's delocalisation in position rather than other factors. This insight proves valuable in deciphering the entanglement generation rate between two bodies prepared in Gaussian or Schrödinger-cat states and subjected to weak interaction forces, such as gravity, a configuration of interest in the context of low-energy tests of the quantum character of gravity.

Often, in the literature, the entanglement rate between two Schrödinger-cat states is computed based solely on the relative dynamical phase build-up. However, we emphasise that this is an approximation. In reality, Schrödinger-cat states interact via gravitational forces in the same manner as any other state, with entanglement arising from the full transformation of the two-particle state under the interaction force. The phase space representation serves as a unifying framework for analysing both cat and Gaussian state configurations, dispelling any notion of privileged configurations for enhanced entanglement capabilities of cat states.

Moreover, our exploration of the phase space representation not only elucidates the equivalence between these configurations but also underscores that not all cat states can be computed in terms of the dynamical phase approach. Squeezed cat states, in particular, necessitate a more refined calculation that considers the full state transformation.

In essence, our paper offers a comprehensive comparison of existing approaches to gravitationally mediated entanglement within the phase-space representation. This framework facilitates a more natural comparison and, in doing so, reveals the origins of their entanglement capabilities. Our work paves the way for a deeper understanding of the intricate interplay between quantum states, weak forces, and entanglement, ultimately advancing the exploration of quantum gravity's enigmatic properties.

\section{Acknowledgements}
We would like to thank Carlo Rovelli and Kirill Streltsov for insightful discussions. This work was supported by the ERC Synergy grant HyperQ (Grant No. 856432)
and the DFG via QuantERA project Lemaqume (Grant no. 500314265). 

With this manuscript, one of us, M.B.P., would like to express gratitude to Professor Sir Peter Knight FRS. Peter not only dedicated 30 years of his service as the Editor of Contemporary Physics but also generously offered his support throughout the same period at critical junctures of M.B.P.'s academic career.

\end{document}